# On a Semi-symmetric Non-metric Connection Satisfying the Schur`s Theorem on a Riemannian Manifold

Ho Tal Yun

Faculty of Mathematics, Kim Il Sung University, D.P.R.K

**Abstract**: in 1992, Agache and Chaple introduced the concept of a semi-symmetric non-metric connection([1]). The semi-symmetric non-metric connection does not satisfy the Schur`s theorem. The purpose of the present paper is to study some properties of a new semi-symmetric non-metric connection satisfying the Schur`s theorem in a Riemannian manifold. And we considered necessary and sufficient condition that a Riemannian manifold with a semi-symmetric non-metric connection be a Riemannian manifold with constant curvature.

**Key words**: semi-symmetric non-metric connection, constant curvature.

## Introduction

In the preceding paper a semi-symmetric non-metric connection was studied in the Riemannian manifold but in the Riemannian manifold with the semi-symmetric non-metric connection Schur's theorem is not proved. In [3] the statistical manifold with constant curvature was studied and in [4] that a statistical manifold with constant curvature is a projective flat manifold was proved. In [5] Schur's theorem was proved in the Riemannin manifold with Levi-Civita Connection.

## Main results of the paper

In this paper we study the semi-symmetric non-metric connection $\nabla$ on a Riemannian manifold $(M, g)$ that satisfies

$$\nabla_Z g(X,Y) = -2\pi(X)g(Y,Z) - 2\pi(Y)g(X,Z)$$
$$T(X,Y) = \pi(Y)X - \pi(X)Y \tag{1}$$

The relation between the semi-symmetric non-metric connection $\nabla$ and the Levi-Civita connection $\overset{\circ}{\nabla}$ of $(M,g)$ is given by

$$\nabla_X Y = \overset{\circ}{\nabla}_X Y + \pi(Y)X + g(X,Y)P \tag{2}$$

where $\pi$ is a 1-form and $P$ is a vector field defined by $g(X,P) = \pi(X)$.

The dual connection $\overset{*}{\nabla}$ of the connection $\nabla$ satisfies on a Riemannian manifold

$$\overset{*}{\nabla}_Z g(X,Y) = 2\pi(X)g(Y,Z) + 2\pi(Y)g(X,Z) \tag{3}$$
$$T(X,Y) = \pi(Y)X - \pi(X)Y$$

The relation between the connection $\overset{*}{\nabla}$ and Levi-Civita connection $\overset{\circ}{\nabla}$ is given by

$$\overset{*}{\nabla}_X Y = \overset{\circ}{\nabla}_X Y - \pi(Y)X - g(X,Y)P \tag{4}$$

If the local expression of $g, \overset{\circ}{\nabla}, \nabla, \overset{*}{\nabla}$ is $g_{ij}, \{^k_{ij}\}, \Gamma^k_{ij}, \overset{*}{\Gamma}^k_{ij}$ respectively, then the local expression of (1), (2), (3) and (4), respectively

$$\nabla_k g_{ij} = -2\varphi_i g_{kj} - 2\varphi_j g_{ki}, \quad T^k_{ij} = \varphi_j \delta^k_i - \varphi_i \delta^k_j \tag{5}$$

$$\Gamma^k_{ij} = \{^k_{ij}\} + \delta^k_i \varphi_j + g_{ij} \varphi^k \tag{6}$$

$$\overset{*}{\nabla}_k g_{ij} = 2\varphi_i g_{kj} + 2\varphi_j g_{ki}, \quad \overset{*}{T}^k_{ij} = \varphi_j \delta^k_i - \varphi_i \delta^k_j \tag{7}$$

$$\overset{*}{\Gamma}^k_{ij} = \{^k_{ij}\} - \delta^k_i \varphi_j - g_{ij} \varphi^k \tag{8}$$

Let $(M, g, \pi)$ be a Riemannian manifold with semi-symmetric non-metric connection $\nabla$.

**Theorem 1**. Suppose $M$ is a connected $n$–dimensional Riemannian manifold with a semi-symmetric non-metric connection $\nabla$ that is everywhere wandering. If $n \geq 3$, then $M$ is a constant curvature space.

Proof. By the second Bianchi identify of a curvature tensor on a Riemannian manifold with an asymmetric non-metric connection

$$\nabla_h R_{ijk}{}^l + \nabla_i R_{hik}{}^l + \nabla_j R_{hik}{}^l = T^m_{hi} R_{jmk}{}^l + T^m_{ij} R_{hmk}{}^l + T^m_{jh} R_{imk}{}^l$$

we obtain

$$\nabla_h R_{ijk}{}^l + \nabla_i R_{hik}{}^l + \nabla_j R_{hik}{}^l = 2(\varphi_h R_{ijk}{}^l + \varphi_i R_{jhk}{}^l + \varphi_j R_{hik}{}^l)$$

If a sectional curvature $k(p)$ at the point $p$ is independent of $E$ (a 2-dimensional subspace of $T_p(M)$), then the curvature tensor is

$$R_{ijk}{}^l = k(p)(\delta_i^l g_{jk} - \delta_j^l g_{ik}).$$ Hence

$$\nabla_h k(\delta_i^l g_{jk} - \delta_j^l g_{ik}) + \nabla_i k(\delta_j^l g_{hk} - \delta_h^l g_{jk}) + \nabla_j k(\delta_h^l g_{ik} - \delta_i^l g_{hk}) +$$
$$+ 2k[\varphi_h(\delta_i^l g_{jk} - \delta_j^l g_{ik}) + \varphi_i(\delta_j^l g_{hk} - \delta_h^l g_{jk}) + \varphi_j(\delta_h^l g_{ik} - \delta_i^l g_{hk})$$
$$= 2k[\varphi_h(\delta_i^l g_{jk} - \delta_j^l g_{ik}) + \varphi_i(\delta_j^l g_{hk} - \delta_h^l g_{jk}) + \varphi_j(\delta_h^l g_{ik} - \delta_i^l g_{hk})$$

From this we obtain
$$\nabla_h k(\delta_i^l g_{jk} - \delta_j^l g_{ik}) + \nabla_i k(\delta_j^l g_{hk} - \delta_h^l g_{jk}) + \nabla_j k(\delta_h^l g_{ik} - \delta_i^l g_{hk}) = 0$$

By contracting with $g^{jk}$, we obtain $(n-2)(\delta_i^l \nabla_h k - \delta_h^l \nabla_i k) = 0$. Contracting indexes $i$ and $l$, we obtain $(n-1)(n-2)\nabla_h k = 0$. Consequently, from $n \geq 3$ we obtain $\nabla_h k = \partial_h k = 0$, that is $k = const$.

**Corollary**. Suppose $M$ is connected $n$-dimensional Riemannian manifold with a dual connection $\overset{*}{\nabla}$ that is everywhere wandering. If $n \geq 3$, then $M$ is a constant curvature space.

By using (6), (8) curvature tensor of $\nabla$ and $\overset{*}{\nabla}$ are respectively

$$R_{ijk}{}^l = K_{ijk}{}^l + \delta_j^l \overset{\circ}{\nabla}_i \varphi_k - \delta_i^l \overset{\circ}{\nabla}_j \varphi_k + g_{jk} \overset{\circ}{\nabla}_i \varphi^l - g_{ik} \overset{\circ}{\nabla}_j \varphi^l \qquad (9)$$
$$- \delta_j^l \varphi_i \varphi_k + \delta_i^l \varphi_j \varphi_k + g_{jk} \varphi_i \varphi^l - g_{ik} \varphi_j \varphi^l + \delta_i^l g_{jk} \varphi_p \varphi^p - \delta_j^l g_{ik} \varphi_p \varphi^p$$

$$\overset{*}{R}_{ijk}{}^l = K_{ijk}{}^l - \delta_j^l \overset{\circ}{\nabla}_i \varphi_k + \delta_i^l \overset{\circ}{\nabla}_j \varphi_k - g_{jk} \overset{\circ}{\nabla}_i \varphi^l + g_{ik} \overset{\circ}{\nabla}_j \varphi^l \qquad (10)$$
$$- \delta_j^l \varphi_i \varphi_k + \delta_i^l \varphi_j \varphi_k + g_{jk} \varphi_i \varphi^l - g_{ik} \varphi_j \varphi^l + \delta_i^l g_{jk} \varphi_p \varphi^p - \delta_j^l g_{ik} \varphi_p \varphi^p$$

where $K_{ijk}{}^l$ is the curvature tensor of $\overset{\circ}{\nabla}$. From (9) and (10) we obtain

$$\overset{*}{R}_{ijk}{}^l = R_{ijk}{}^l + 2(\delta_i^l \overset{\circ}{\nabla}_j \varphi_k - \delta_j^l \overset{\circ}{\nabla}_i \varphi_k + g_{ik} \overset{\circ}{\nabla}_j \varphi^l - g_{jk} \overset{\circ}{\nabla}_i \varphi^l) \qquad (11)$$

**Lemma 1.** If the Weyl conformal curvature of $\overset{\circ}{\nabla}, \nabla$ and $\overset{*}{\nabla}$ are $\overset{\circ}{C}_{ijk}{}^l$, $C_{ijk}{}^l$ and $\overset{*}{C}_{ijk}{}^l$, respectively, then

$$C_{ijk}{}^l + \overset{*}{C}_{ijk}{}^l = 2\overset{\circ}{C}_{ijk}{}^l \qquad (12)$$

Proof. From (9) and (10), we have

$$R_{ijk}{}^l + \overset{*}{R}_{ijk}{}^l = 2(K_{ijk}{}^l + \delta_i^l \varphi_{jk} - \delta_j^l \varphi_{ik} + g_{jk} \varphi_i^l - g_{ik} \varphi_j^l) \qquad (13)$$

where $\varphi_{jk} = \varphi_j \varphi_k + \frac{1}{2} g_{jk} \varphi_p \varphi^p$, contracting indexes $i$ and $l$

$$R_{jk} + \overset{*}{R}_{jk} = 2(K_{jk} + (n-2)\varphi_{jk} + g_{jk}\varphi_i^i) \tag{14}$$

By contracting with $g^{jk}$

$$\varphi_i^i = \frac{1}{4(n-1)}(R + \overset{*}{R} - 2k) \tag{*}$$

Substituting (*) into (14)

$$\varphi_{jk} = \frac{1}{2(n-2)}[R_{jk} + \overset{*}{R}_{jk} - 2K_{jk} - \frac{R + \overset{*}{R} - 2k}{2(n-1)}] \tag{15}$$

Also substituting (15) into (13), we have (12). Thus we have:

**Theorem 2.** If a Riemannian metric admits a semi-symmetric non-metric connection $\nabla$ whose curvature tensor and dual curvature tensor vanishes on a Riemannian manifold, then the Riemannian metric is conformal flat.

**Theorem 3.** In order that a Riemannian metric with a constant curvature is admitted on $(M,g,\pi)$, it is necessary and sufficient that a semi-symmetric non-metric connection $\nabla$ should be a conjugate symmetric and a conformal flat connection, and that its Ricci curvature tensor satisfy the Einstein equation

Proof. From $R_{ijk}^{\ l} = k(\delta_i^l g_{jk} - \delta_j^l g_{ik})$, $R_{ijkl} = k(g_{il}g_{jk} - g_{jk}g_{ik})$, we obtain

$$\overset{*}{R}_{ijkl} = -R_{ijlk} = k(g_{ik}g_{jl} - g_{jk}g_{il}) = k(g_{il}g_{jk} - g_{ik}g_{jl}) = R_{ijkl}$$

Thus $\nabla$ is a conjugate symmetric connection. On the other hand, from $R_{jk} = \overset{*}{R}_{jk} = k(n-1)g_{jk}$, Ricci curvature $R_{jk}$ satisfies the Einstein equation. And from $R = \overset{*}{R} = k(n-1)$ and lemma 1, $C_{ijk}^{\ l} = \overset{*}{C}_{ijk}^{\ l} = 0$. Thus $g_{ij}$ is a conformal flat metric.

Conversely, if $\nabla$ is a conjugate symmetric and a conformal flat connection, then from lemma 1

$$R_{ijk}^{\ l} = \frac{1}{n-1}(\delta_i^l R_{jk} - \delta_j^l R_{ik} + g_{jk}R_i^l - g_{ik}R_j^l) + \frac{R}{(n-1)(n-2)}(\delta_j^l g_{ik} - \delta_i^l g_{jk})$$

Using the Einstain equation $R_{ij} = \frac{R}{n}g_{ij}$, we find $R_{ijk}^{\ l} = \frac{R}{n(n-1)}(\delta_i^l g_{jk} - \delta_j^l g_{ik})$. Thus the Riemannian metric is of constant curvature on $(M, g, \pi)$.

**Lemma 2**. The curvature tensor $R_{ijk}{}^l$ of a Riemannian manifold with a semi-symmetric non-metric connection $\nabla$ satisfies the following properties:

1) $R_{(ijk)}{}^l + \overset{*}{R}_{(ijk)}{}^l = 0$

2) $R_{jk} + \overset{*}{R}_{jk} = R_{kj} + \overset{*}{R}_{kj}$

3) If 1-form $\pi$ is closed, then $R_{(ijk)}{}^l = 0$, $R_{jk} = R_{kj}$, $P_{ij} = 0$

$\overset{*}{R}_{(ijk)}{}^l = 0$, $\overset{*}{R}_{jk} = \overset{*}{R}_{kj}$, $\overset{*}{P}_{ij} = 0$

where ( ) represents circle permutation and $P_{ij}$, $\overset{*}{P}_{ij}$ are the volume curvature tensor of $\nabla$, $\overset{*}{\nabla}$.

Proof. From (9) and (10), we obtain

$R_{(ijk)}{}^l = \delta_i^l(\overset{\circ}{\nabla}_k \varphi_j - \overset{\circ}{\nabla}_j \varphi_k) + \delta_j^l(\overset{\circ}{\nabla}_i \varphi_k - \overset{\circ}{\nabla}_k \varphi_i) + \delta_k^l(\overset{\circ}{\nabla}_j \varphi_i - \overset{\circ}{\nabla}_i \varphi_j)$

$\overset{*}{R}_{(ijk)}{}^l = \delta_i^l(\overset{\circ}{\nabla}_j \varphi_k - \overset{\circ}{\nabla}_k \varphi_j) + \delta_j^l(\overset{\circ}{\nabla}_k \varphi_i - \overset{\circ}{\nabla}_i \varphi_k) + \delta_k^l(\overset{\circ}{\nabla}_i \varphi_j - \overset{\circ}{\nabla}_j \varphi_i)$

By using these, we prove Lemma 2.

Now we consider connection transformations according to conformal transformation of the metric in a Riemannian manifold (M, g):

$g_{ij} \to \bar{g}_{ij} = e^{2\sigma} g_{ij}$.

Corresponding connection transformations $\nabla \to \bar{\nabla}$ and $\overset{*}{\nabla} \to \overset{\bar{*}}{\nabla}$ are represented locally respectively;

$$\bar{\Gamma}_{ij}^k = \Gamma_{ij}^k + \sigma_i \delta_j^k + \sigma_j \delta_i^k - g_{ij} \sigma^k$$
$$\overset{*}{\bar{\Gamma}}_{ij}^k = \overset{*}{\Gamma}_{ij}^k + \sigma_i \delta_j^k + \sigma_j \delta_i^k - g_{ij} \sigma^k \qquad (16)$$

where $\sigma_i = \dfrac{\partial \sigma}{\partial x^i}$.

**Theorem 4.** If the tensor $U_{ijk}{}^l$ is given by $U_{ijk}{}^l = R_{ijk}{}^l + \overset{*}{R}_{ijk}{}^l$, then the invariant of the connection transformation (16) is the Weyl conformal curvature tensor of $\bar{U}_{ijk}{}^l$, that is

$$\overset{U}{\overline{C}}_{ijk}{}^l = \overline{U}_{ijk}{}^l + \frac{1}{n-2}(\delta_j^l \overline{U}_{ik} - \delta_i^l \overline{U}_{jk} + g_{ik}\overline{U}_j^l - g_{jk}\overline{U}_i^l)$$
$$+ \frac{\overline{U}}{(n-1)(n-2)}(\delta_i^l g_{jk} - \delta_j^l g_{ik}) \quad (17)$$

Proof. From (16) the curvature tensors are

$$\overline{R}_{ijk}{}^l = R_{ijk}{}^l + \delta_j^l \overset{\circ}{\nabla}_i \sigma_k - \delta_i^l \overset{\circ}{\nabla}_j \sigma_k + g_{ik} \overset{\circ}{\nabla}_j \sigma^l - g_{jk} \overset{\circ}{\nabla}_i \sigma^l + \delta_i^l(\sigma_j \varphi_k + \sigma_k \varphi_j)$$
$$- \delta_j^l(\sigma_i \varphi_k + \sigma_k \varphi_i) + g_{ik}(\varphi_j \sigma^l + \sigma_j \varphi^l) - g_{jk}(\varphi_i \sigma^l + \sigma_i \varphi^l) + \delta_i^l \sigma_j \sigma_k$$
$$- \delta_j^l \sigma_i \sigma_k + g_{ik}\sigma_j \sigma^l - g_{jk}\sigma_i \sigma^l + \delta_j^l g_{ik}\sigma_p \sigma^p - \delta_i^l g_{jk}\sigma_p \sigma^p$$

$$\overset{*}{R}_{ijk}{}^l = R_{ijk}{}^l + \delta_j^l \overset{\circ}{\nabla}_i \sigma_k - \delta_i^l \overset{\circ}{\nabla}_j \sigma_k + g_{ik} \overset{\circ}{\nabla}_j \sigma^l - g_{jk} \overset{\circ}{\nabla}_i \sigma^l + \delta_i^l(\sigma_j \varphi_k + \sigma_k \varphi_j)$$
$$- \delta_j^l(\sigma_i \varphi_k + \sigma_k \varphi_i) + g_{ik}(\varphi_j \sigma^l + \sigma_j \varphi^l) - g_{jk}(\varphi_i \sigma^l + \sigma_i \varphi^l) + \delta_i^l \sigma_j \sigma_k - \delta_j^l \sigma_i \sigma_k$$
$$+ g_{ik}\sigma_j \sigma^l - g_{jk}\sigma_i \sigma^l + \delta_j^l g_{ik}\sigma_p \sigma^p - \delta_i^l g_{jk}\sigma_p \sigma^p$$

By summing these equations we have
$$\overline{U}_{ijk}{}^l = U_{ijk}{}^l + \delta_j^l \sigma_{ik} - \delta_i^l \sigma_{jk} + g_{ik}\sigma_j^l - g_{jk}\delta_i^l \quad (18)$$

where $\sigma_{ik} = 2(\overset{\circ}{\nabla}_i \delta_k - \sigma_i \sigma_k + \frac{1}{2}g_{jk}\sigma_i^i)$. Contracting indexes $i$, $l$ in (18)

$$\overline{U}_{jk} = U_{jk} + (2-n)\sigma_{jk} - g_{jk}\sigma_i^i \quad (*)$$

Also contracting with $g^{jk}$, $\overline{U} = U + 2(n-1)\sigma_i^i$. Substituting $\sigma_i^i$ into (*), we get $\sigma_{jk} = \frac{1}{n-2}(\overline{U}_{jk} - U_{jk} - \frac{\overline{U}-U}{2(n-1)}g_{jk})$. And substituting $\sigma_{jk}$ into (18) and putting

$$\overset{U}{C}_{ijk}{}^l = U_{ijk}{}^l + \frac{1}{n-2}(\delta_j^l U_{ik} - \delta_i^l U_{jk} + g_{ik}U_j^l - g_{jk}U_i^l) + \frac{U}{(n-1)(n-2)}(\delta_i^l g_{jk} - \delta_j^l g_{ik})$$

we obtain $\overset{U}{\overline{C}}_{ijk}{}^l = \overset{U}{C}_{ijk}{}^l$. Thus we have:

**Corollary 1.** If the semi-symmetric non-metric connection $\nabla$ is a conjugate symmetric connection (that is $R_{ijk}{}^l = \overset{*}{R}_{ijk}{}^l$), then the invariant of the connection transformation (16) is the Weyl conformal curvature tensor of the curvature tensor $\overline{R}_{ijk}{}^l$.

**Corollary 2.** If the semi-symmetric non-metric connection $\nabla$ is a conjugate symmetric connection (that is $R_{ijk}{}^l = \overset{*}{R}_{ijk}{}^l$), then the invariant of the connection transformations (6), (8) is the Weyl conformal curvature tensor of the curvature tensor $R_{ijk}{}^l$.

**References**

[1] Agache Nimala S, Chafle Mangala R, A semi-symmetric non-metric connection on a Riemannian manifold. Indian J. Pure. Appl. Math. 23(1992) No 6 399-409.
[2] De. U. C, Biswas. S. C, On a type of semi-symmetric non-metric connection on a Riemannian manifold. Istanbul Univ. Fen Fak. Mat. Derg. 55/56 (1996/97) 237-243.
[3] Kurose T and eto al, On the divergences of 1-conformally flat statistical manifolds. Tohoku Math.J.(2), 4b,427-433,1994
[4] E. S. Stepanova, Dual Symmetric statistical manifolds. Journal of Mathematical Science,vol,147, No1, 6507-6509,2007
[5] S. S. Chern, W. H. Chen and K. S. Lam, Lectures on Differential Geometry, World Scientific, 2000, 160.